%% file: main.tex
\documentclass[a4paper,12pt]{article}

\usepackage{jheppub}

\usepackage{amsmath,amssymb,amsfonts}

\usepackage{graphicx}

\usepackage{subcaption}

\usepackage{slashed}      
\usepackage{bm}           

\usepackage{hyperref}

\usepackage{cite}
\title{Sensitivity to Higgs Pseudo-Observables in ZH Production at NLO QCD}

\author[a]{Makhlouf Chennit}
\author[b]{Abdelkader \text{Mohamed-Meziani}}

\affiliation[a]{Laboratory of Théoretical physics, University of  Bejaia, \text{Bejaia-06000- Algeria}}
\affiliation[b]{ Departement of physics, Faculty of Exactes Sciences, University of Bejaia, \text{Bejaia 06000- Algeria}}

\emailAdd{makhlouf.chennit@univ-bejaia.dz}

\emailAdd{abdelkader.mohamedmeziani@univ-bejaia.dz}

\abstract{We study the sensitivity of the associated Higgs
production process $pp \rightarrow ZH$ to anomalous
Higgs couplings within the Effective Field Theory
framework. Event samples are generated using
$ \text{MadGraph5\_aMC@NLO}$ at $\sqrt{s}=13$ TeV using the UFO model HPO\_ewk\_prod\_NLO, and analyzed
using a dedicated Python analysis of HepMC events.
We investigate several kinematic observables and
derive constraints on the EFT parameter
$\epsilon_{ZZ}$ using a $\chi^2$ statistical analysis.
A 95\% confidence level interval
 $-0.026 < \epsilon_{ZZ} < 0.383 $
is obtained from the simulated data.}

\begin{document}
\maketitle
\flushbottom
\newpage

\input{sections/Introduction}

\input{sections/EFT_and_HPO_Framework}

\input{sections/results}

\input{sections/Conclusion}


\bibliographystyle{JHEP}
\bibliography{references}

\end{document}

%% file: sections/Introduction.tex
\section{Introduction}

The discovery of the Higgs boson in the year of 2012, by the
experiments at the Large Hadron Collider (LHC),
is one of the most important achievements
in particle physics.
This discovery confirmed the mechanism of electroweak
symmetry breaking \cite{Englert:1964et,Higgs:1964pj} predicted by the Standard Model (SM) \cite{Glashow:1961tr,Weinberg:1967tq,Salam:1968rm}
and completed the particle content of the theory.

The measure  of the Higgs boson properties
are consistent with the predictions of the
Standard Model, the measurement of its
interactions are the main objective of the LHC
physics program.
In particular, deviations from the SM expectations
may provide indirect evidence for new physics
beyond the Standard Model.

One of the most interesting approaches to study possible
new physics effects in the Higgs sector is the
Effective Field Theory (EFT) framework \cite{eft_review}.
Within this approach, the Standard Model Lagrangian
contains the higher-dimensional operators that
carries the effects of heavy new particles at an
energy scale $\Lambda$, larger than the
electroweak scale.
These operators modify the couplings of the Higgs
boson, witch affect both production
cross sections and kinematic distributions.

For the different Higgs production mechanisms, the associated production of the Higgs
boson with a vector boson plays an
important role.\cite{atlas_higgs,cms_higgs}

The process
$
pp \rightarrow ZH
$
 called Higgsstrahlung specifies direct sensitivity to the Higgs coupling
to electroweak gauge bosons.
This process presents a clean experimental signature
and has been studied by the ATLAS
and CMS collaborations.\cite{atlas_higgs,cms_higgs}

The production rate and kinematic properties of
the $ZH$ process are directly related to the
structure of the $HZZ$ interaction.
The deviations from the Standard Model
coupling can  be explored through precise
measurements of this channel.

In the absence of direct discoveries of new particles,
precision measurements of Higgs boson interactions
teach us the ways to explore physics beyond the Standard Model.
In this context, the EFT framework allows a
systematic parameterization of potential deviations
from the Standard Model predictions.

In this work we investigate the sensitivity of
associated Higgs production to anomalous Higgs
couplings within the EFT framework.
We focus in particular on the parameter
$\epsilon_{ZZ}$ which modifies the interaction
between the Higgs boson and the $Z$ boson.

Event samples are generated at next-to-leading order
using $ \texttt{MadGraph5\_aMC@NLO}$ \cite{mg5} within the HPO model \cite{Gonzalez-Alonso:2014eva,Greljo:2015sla}.
The process
$pp \rightarrow ZH$
is simulated at a center-of-mass energy of
$
\sqrt{s}=13\ \text{TeV}.
$

The invariant mass of the $ZH$ system and several kinematic observables are analyzed,
including the transverse momentum of the Higgs
boson and the Z boson.

A statistical analysis based on the total cross
section is then performed in order to derive
constraints on the EFT parameter $\epsilon_{ZZ}$.

The structure of this paper is as follows. In Section 2 we briefly review Effec-
tive Field Theory and Higgs Pseudo-Observables Framework used in this analysis.
Section 3 describes the implemention in monte carlo simulation. The results and
discussion are presented in Section 4. Finally, the conclusions are summarized in the
last section 5.

%% file: sections/EFT_and_HPO_Framework.tex
\section{Effective Field Theory and Higgs Pseudo-Observables \\ Framework}
\indent

The Standard Model (SM) of particle physics \cite{Glashow:1961tr,Weinberg:1967tq,Salam:1968rm} provides a
remarkably successful description of the interactions
between elementary particles. However, it is widely
believed to be an effective theory valid up to a certain
energy scale, beyond which new physical phenomena may
emerge.

In the absence of direct evidence for new particles, the
Effective Field Theory (EFT) \cite{eft_review} approach offers a powerful
and model-independent framework to describe possible
deviations from the Standard Model predictions. In this
framework, the SM Lagrangian is extended by higher–
dimensional operators that encode the effects of heavy
new physics at an energy scale $\Lambda$.

The EFT Lagrangian can be written as \cite{BrivioTrott2019}

\begin{equation}
\mathcal{L}_{\mathrm{EFT}} =
\mathcal{L}_{\mathrm{SM}} +
\sum_i \frac{c_i}{\Lambda^2} \mathcal{O}_i +
\mathcal{O}\left(\frac{1}{\Lambda^4}\right),
\end{equation}

where $\mathcal{O}_i$ are dimension–six operators and
$c_i$ are the corresponding Wilson coefficients.

\vspace{0.3cm}


The Higgs pseudo-observables (HPO) framework \cite{Gonzalez-Alonso:2014eva,Greljo:2015sla} provides a
complementary and experimentally oriented parameterization
of possible deviations in Higgs boson interactions. It
allows one to directly relate measurable quantities at
colliders to effective couplings, without relying on a
specific ultraviolet completion of the theory.

In this approach, the effects of new physics are encoded
in a set of effective couplings that modify both the
strength and the Lorentz structure of Higgs interactions.
These parameters can be matched to the coefficients of
the EFT operators, establishing a direct connection
between the two formalisms.

\vspace{0.3cm}


In the HPO framework, deviations from the SM are commonly
parameterized through effective couplings such as
$
\kappa_{ZZ}, \quad \kappa_{WW}, \quad
\kappa_{\gamma\gamma}, \quad \kappa_{Z\gamma},
$
as well as momentum-dependent parameters like
$\epsilon_{ZZ}$.

A simplified effective Lagrangian describing the
interaction between the Higgs boson and the $Z$ boson
can be written as  \cite{GonzalezAlonso2015}
\begin{equation}
\mathcal{L}_{HZZ} =
\kappa_{ZZ}\, g_{HZZ}^{\mathrm{SM}}\, H Z_\mu Z^\mu
+ \frac{\epsilon_{ZZ}}{\Lambda^2}\,
H Z_{\mu\nu} Z^{\mu\nu},
\end{equation}

where $Z_{\mu\nu}$ denotes the field strength tensor
of the $Z$ boson.

The first term corresponds to a rescaling of the Standard
Model coupling, while the second term introduces a
momentum-dependent interaction that becomes increasingly
important at high energies.

The most general amplitude for the interaction between
the Higgs boson and two neutral gauge bosons can be
expressed as \cite{Greljo2016}
\begin{equation}
\mathcal{A}(H \rightarrow ZZ) =
\mathcal{A}_{\mathrm{SM}} +
\epsilon_{ZZ}\, \mathcal{A}_{\mathrm{BSM}} + \cdots,
\end{equation}

where $\epsilon_{ZZ}$ parametrizes deviations of the
Standard Model contribution.
\vspace{0.3cm}

\section{Implementation in Monte Carlo Simulation}
\indent

In this study, Monte Carlo event samples are generated to investigate the sensitivity of the process $pp \rightarrow ZH$ to anomalous Higgs couplings within the Higgs Pseudo-Observables (HPO) framework.

The process is simulated for proton-proton collisions at a center of $\sqrt{s}=13$ TeV at LHC. 
The collisions are generated at NLO next to leading order with QCD and QED corrections  
using the \texttt{HPO\_ewk\_prod\_NLO} model\cite{Greljo:2017spw} embedded in the MadGraph5\_aMC@NLO event generator.\\
We use the parton distrbution functions \text{NNPDF23\_nlo\_as\_0119\_qed} \cite{Ball2013QED}, as implemented in LHAPDF set up \cite{lhapdf2015}.
The generated events are obtained from simulations with  \texttt{PYTHIA8} \cite{pythia8} for parton showering and hadronization.

Within this setup, the effective Higgs couplings are parameterized using coupling modifiers. All Standard Model-like couplings are fixed to their nominal values,
\begin{equation}
\kappa_{ZZ} = 1, \quad
\kappa_{WW} = 1, \quad
\kappa_{\gamma\gamma} = 1, \quad
\kappa_{Z\gamma} = 1,
\end{equation}
ensuring that any deviation from the Standard Model originates solely from higher-dimensional operators. In particular, the parameter $\epsilon_{ZZ}$, which introduces momentum-dependent corrections to the HZZ interaction, is varied to probe possible new physics effects.

To explore the sensitivity of observables to anomalous couplings, three benchmark scenarios are defined:
\begin{equation}
\epsilon_{ZZ} = -0.5, \quad 0, \quad 0.5.
\end{equation}

These benchmark points allow for a systematic study of the dependence of both the total production cross section and kinematic distributions on the strength of the EFT contribution.

For each benchmark scenario, a total of $N = 100000$ events are generated. The resulting cross sections exhibit a strong dependence on the value of $\epsilon_{ZZ}$, reflecting the sensitivity of the ZH production channel to modifications of the Higgs coupling to the Z boson.

After parton showering with \texttt{PYTHIA8}, the generated events are stored in HepMC format \cite{hepmc} and are subsequently analyzed using a Python-based library \cite{pyhepmc} within a JupyterLab environment \cite{jupyter2016}. From these samples, several kinematic observables are reconstructed, including the transverse momenta of the Higgs and Z bosons, the invariant mass of the ZH system, and angular correlations.

%% file: sections/results.tex
\section{Results and discussion}
\indent

The generated events for the process $pp \rightarrow ZH$
are stored in the HepMC event record format
and subsequently analyzed using the \texttt{pyhepmc}
Python library. This framework provides
efficient access to particle-level information and enables
a flexible analysis workflow within a Python-based
environment.

The analysis is performed within the JupyterLab
environment, where dedicated Python scripts are used
to process the simulated event samples and extract the
relevant kinematic observables.
For each generated event, the full list of final-state
particles is scanned in order to identify the Higgs
boson and the $Z$ boson present in the event record.
These particles are uniquely identified using their
Particle Data Group (PDG) identifiers
\begin{equation}
\text{PDG}(H) = 25, \qquad \text{PDG}(Z) = 23,
\end{equation}
ensuring a robust and unambiguous selection.
Once the Higgs and $Z$ bosons are identified, their
four--momenta are used to reconstruct a set of
kinematic observables sensitive to anomalous Higgs
couplings.

The transverse momentum of each particle is defined as
\begin{equation}
p_T = \sqrt{p_x^2 + p_y^2},
\end{equation}
where $p_x$ and $p_y$ denote the momentum components
in the plane transverse to the beam axis.
The invariant mass of the $ZH$ system is computed as
\begin{equation}
m_{ZH} =
\sqrt{(E_H + E_Z)^2 - (\vec{p}_H + \vec{p}_Z)^2}.
\end{equation}

The transverse momentum of the combined $ZH$ system
is obtained from the vector sum of the transverse
momenta of the Higgs and $Z$ bosons.
\begin{equation}
p_T^{ZH}
=
\sqrt{
(p_T^Z)^2
+
(p_T^H)^2
+
2\,p_T^Z p_T^H \cos(\Delta\phi_{ZH})
}
\end{equation}

In addition, the azimuthal separation between the
two particles is defined as
\begin{equation}
\Delta\phi(Z,H) = \phi_Z - \phi_H.
\end{equation}

This observable provides information on the event
topology and can be sensitive to modifications of
the underlying interaction.

For each event sample corresponding to a given value
of the EFT parameter $\epsilon_{ZZ}$, the distributions
of the observables we are interested in are constructed.
The distributions are obtained by accumulating the
values of these observables over the full set of
generated events.

A comparison of the resulting distributions for the
different benchmark values of $\epsilon_{ZZ}$ allows
us to probe the impact of anomalous Higgs couplings
on the production kinematics. In particular, deviations
from the Standard Model prediction are expected to
become more pronounced in the high-energy regions of
phase space, where the effects of higher-dimensional
operators are enhanced \cite{Biekoetter2015}.

These observables therefore provide complementary
information to the total production cross section
and play a crucial role in assessing the sensitivity
of the $ZH$ production channel to modifications of
the $HZZ$ interaction.
The resulting total production cross sections obtained
from the event generation are

\begin{table}[ht]
	\centering
	\renewcommand{\arraystretch}{1.5}
	\setlength{\tabcolsep}{0.8cm}
\begin{tabular}{cc}
	\hline
	$\epsilon_{ZZ}$ & $\sigma$ [pb] \\
	\hline
	0.0 & $0.7321 \pm 0.0015$ \\
	0.5 & $1.7270 \pm 0.0033$ \\
	-0.5 & $6.6710 \pm 0.0140$ \\
	\hline
\end{tabular}

	\caption{the total production cross section for the benchmark points}
	\label{Tab:1}
\end{table}

In addition to the total production cross section,
kinematic distributions provide valuable information
about the underlying dynamics of the production
process and offer enhanced sensitivity to possible
deviations from the Standard Model predictions. We analyze several observables
constructed from the four--momenta of the Higgs
boson and the $Z$ boson in the simulated event
samples.

\begin{itemize}

\item transverse momentum of the Higgs boson $p_T(H)$,

\item transverse momentum of the $Z$ boson $p_T(Z)$,

\item invariant mass of the $ZH$ system $m_{ZH}$,

\item transverse momentum of the $ZH$ system $p_T(ZH)$,

\item azimuthal separation between the Higgs boson
and the $Z$ boson $\Delta\phi(Z,H)$.

\end{itemize}
The following kinematic variables are considered.

The distributions corresponding to different values
of the EFT parameter $\epsilon_{ZZ}$ are compared in
order to evaluate the impact of anomalous Higgs
couplings on the event kinematics.
To enhance sensitivity to EFT effects we define a boosted
selection

\begin{equation}
p_T(H) > 300~\text{GeV,}\quad m_{ZH} > 600~\text{GeV} \quad \Delta \phi(Z,H) > 2.5 \quad p_T(ZH) > 200~\text{GeV}
\end{equation}

\subsection*{Transverse Momentum Distribution of the Higgs Boson}
\indent
The transverse momentum of the Higgs boson is one of
the most sensitive observables to modifications of
the Higgs interaction with electroweak gauge bosons.

\begin{figure}[h]
\centering
\includegraphics[width=0.7\textwidth]{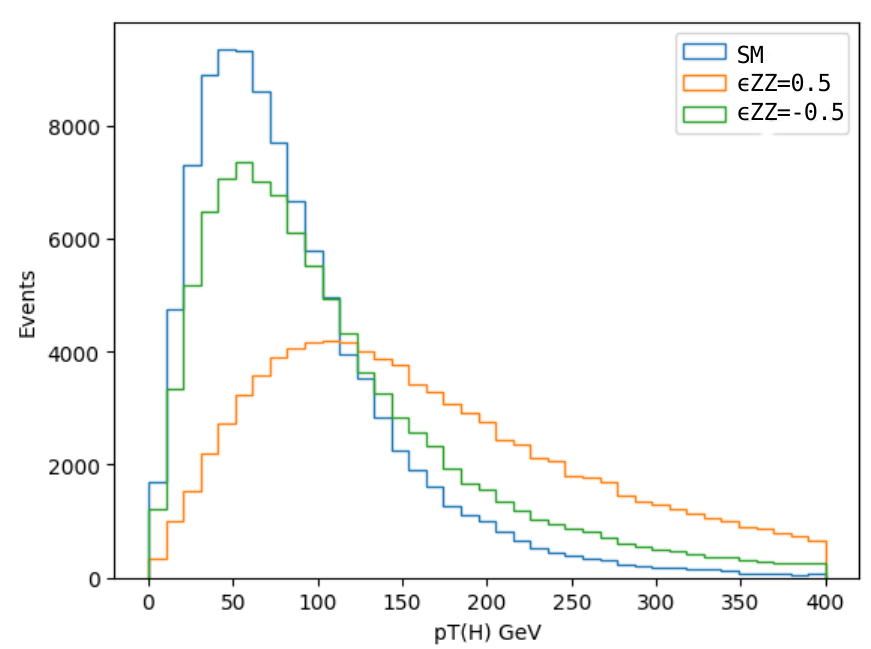}
\caption{Transverse momentum distribution of the Higgs boson
for different values of the EFT parameter $\epsilon_{ZZ}$.}
\label{fig:ptH}
\end{figure}

Figure~1 shows the transverse momentum distributions of the Higgs boson $p_T(H)$ for the Standard Model (SM) and two benchmark scenarios corresponding to anomalous couplings $\epsilon_{ZZ} = +0.5$ and $\epsilon_{ZZ} = -0.5$.

The Standard Model prediction exhibits the typical behaviour expected for Higgs production in association with a $Z$ boson. The distribution peaks around moderate transverse momentum values ($\sim 50$--$70~\mathrm{GeV}$) and then rapidly decreases toward higher $p_T$.

When anomalous couplings are introduced, noticeable modifications of the spectrum appear. In particular, the scenario with $\epsilon_{ZZ}=+0.5$ leads to a significant distortion of the distribution. The peak is shifted toward higher transverse momentum and the high-$p_T$ tail becomes substantially enhanced compared to the SM prediction. This results means that more Higgs bosons are produced with high transverse momentum $(p_{T}^{H})$ than expected in the Standard Model (SM).

The benchmark point with $\epsilon_{ZZ}=-0.5$ also modifies the distributions with respect to the Standard Model, the effect is less pronounced than for the positive coupling scenario. The spectrum remains closer to the SM shape but still shows a moderate enhancement in the intermediate and high-$p_T$ regions.

This behaviour is consistent with expectations from the Effective Field Theory framework, where anomalous Higgs--gauge boson interactions tend to produce harder kinematic spectra. Consequently, the transverse momentum distribution of the Higgs boson provides a sensitive observable to probe deviations from the Standard Model and constrain the anomalous couplings parameter $\epsilon_{ZZ}$.

\subsection*{Transverse Momentum of the $Z$ Boson}
The transverse momentum of the $Z$ boson provides
complementary information about the production
dynamics of the associated Higgs process.

\begin{figure}[h]
\centering
\includegraphics[width=0.7\textwidth]{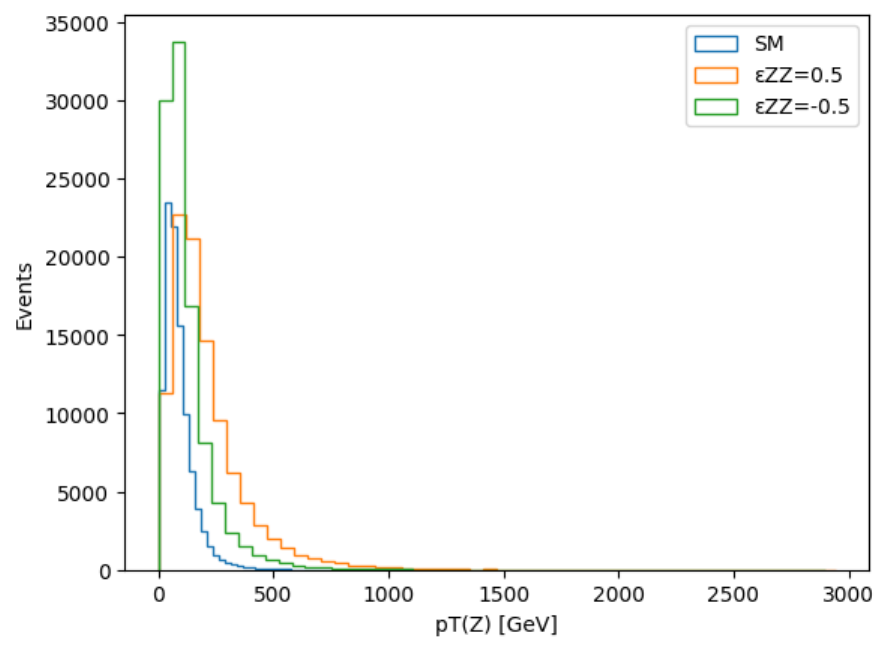}
\caption{Transverse momentum distributions of the
$Z$ boson.}
\label{fig:ptZ}
\end{figure}

Figure~2 shows the transverse momentum distributions of the $Z$ boson, $p_T(Z)$, is shown for the Standard Model (SM) and for two benchmark scenarios with anomalous couplings $\epsilon_{ZZ}=0.5$ and $\epsilon_{ZZ}=-0.5$. One observes that the SM prediction dominates in the low transverse momentum region, where the three distributions exhibit similar behaviors. However, significant deviations appear as the transverse momentum increases. In particular, the scenario with $\epsilon_{ZZ}=0.5$ shows a noticeable enhancement in the high-$p_T$ tail compared to the SM prediction, while the $\epsilon_{ZZ}=-0.5$ case leads to a suppression of events. This behavior is consistent with the expectation that contributions from higher-dimensional operators in the Effective Field Theory framework become more pronounced at high energies. Consequently, the high-$p_T(Z)$ region provides an important probe for potential deviations from the Standard Model and offers sensitivity to new physics effects in the $HZZ$ interaction.

Since the Higgs and $Z$ bosons are produced together,
their transverse momenta are strongly correlated.
Consequently, modifications of the Higgs coupling
also affect the kinematic properties of the $Z$
boson.

\subsection*{Invariant Mass of the $ZH$ System}

Another important observable is the invariant mass
of the $ZH$ system
 
\begin{figure}[h!]
	\centering
	\begin{subfigure}{0.52\textwidth}
		\centering
		\includegraphics[width=\linewidth, height=5cm, keepaspectratio=false]{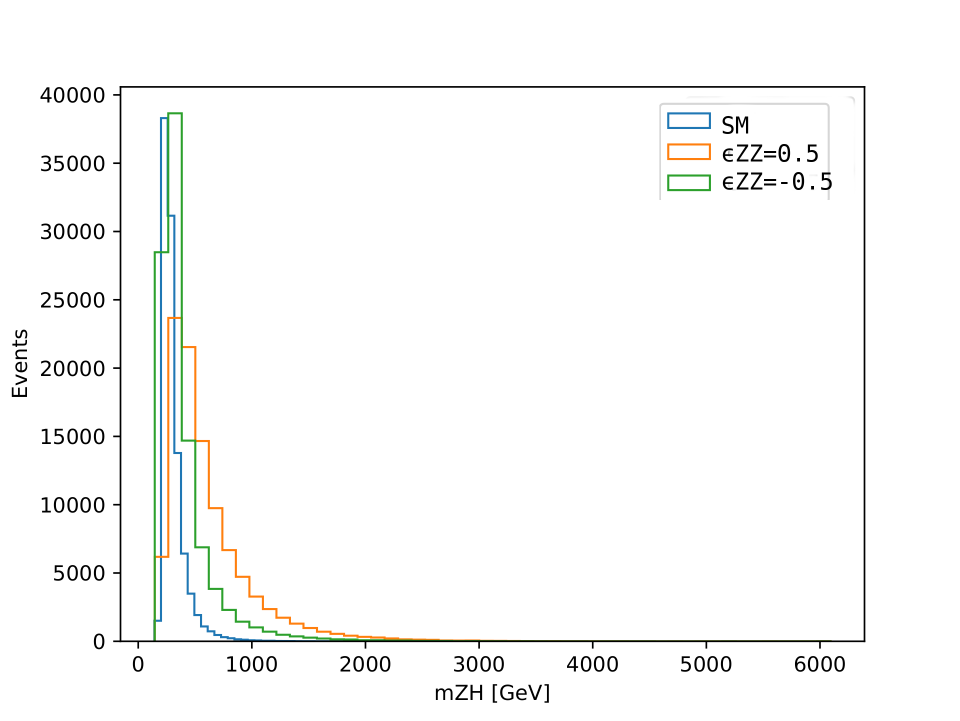}
		\caption{$m(ZH)$}
	\end{subfigure}
	\hfill
	\begin{subfigure}{0.45\textwidth}
		\centering
		\includegraphics[width=\linewidth, height=5cm, keepaspectratio=false]{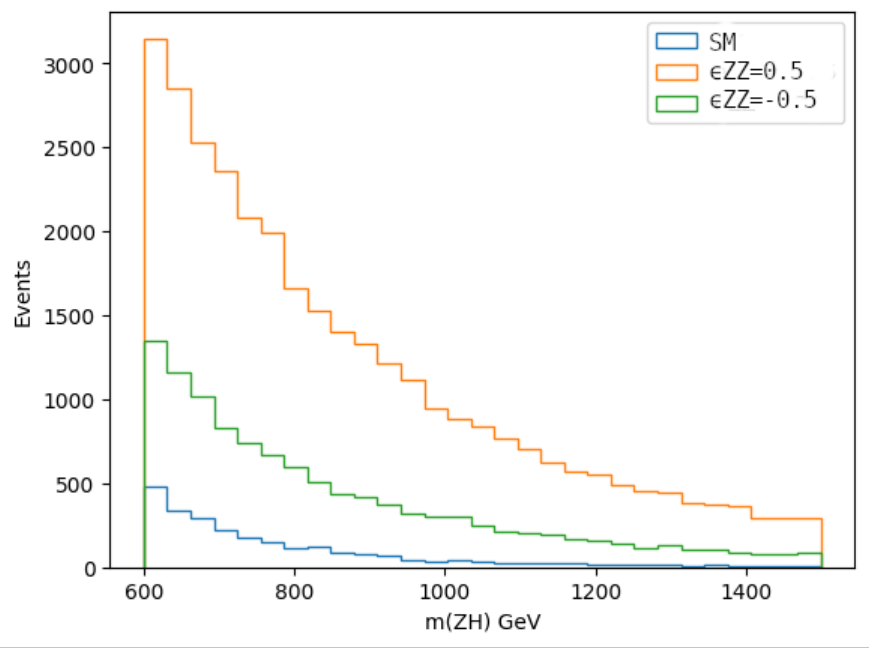}
		\caption{$m(ZH) > 600$}
	\end{subfigure}
	\caption{distribution $m(ZH)$ (a) and distribution $m(ZH)$ in boosted region (b)}
\end{figure}

Figure~3 shows the invariant mass distribution of the $ZH$ system, $m(ZH)$. It is presented for the Standard Model (SM) and for two benchmark scenarios with anomalous couplings $\epsilon_{ZZ}=0.5$ and $\epsilon_{ZZ}=-0.5$. In the low invariant mass region, the SM prediction dominates and the three scenarios show relatively similar behaviors. As the invariant mass increases, significant deviations from the SM prediction become visible. In particular, the scenario with $\epsilon_{ZZ}=0.5$ exhibits a strong enhancement in the high-mass tail, indicating the presence of harder kinematic configurations. The $\epsilon_{ZZ}=-0.5$ scenario also shows an excess compared to the SM, but less pronounced. These effects are expected in the Effective Field Theory framework, where contributions from higher-dimensional operators grow with the energy scale of the process. Therefore, the high-$m(ZH)$ region provides an important window to probe anomalous $HZZ$ interactions and potential deviations from the Standard Model predictions 

\subsection*{Transverse Momentum Distribution of the $ZH$ System}

The transverse momentum of the combined $ZH$ system
is obtained from the vector sum of the transverse
momenta of the Higgs and $Z$ bosons.

\begin{figure}[h!]
    \centering
    \begin{subfigure}{0.51\textwidth}
        \centering
        \includegraphics[width=\linewidth]{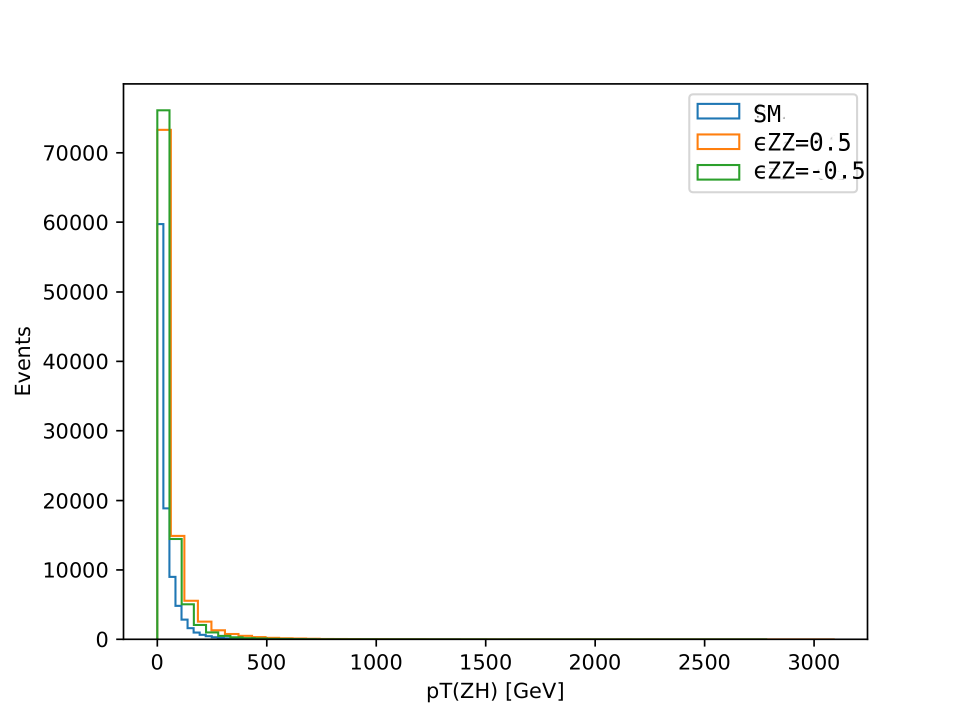}
        \caption{$p_T(ZH)$}
    \end{subfigure}
    \hfill
    \begin{subfigure}{0.45\textwidth}
        \centering
        \includegraphics[width=\linewidth]{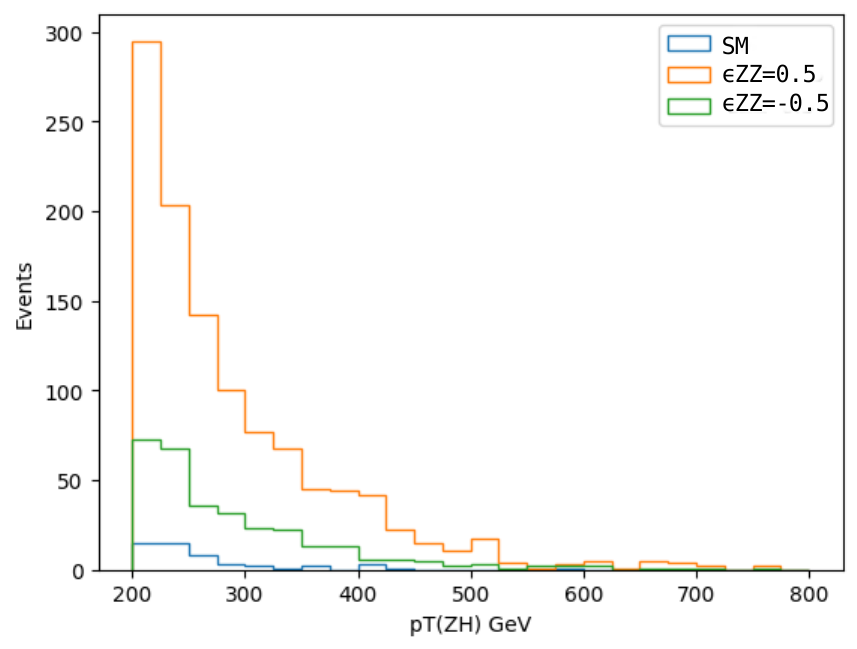}
        \caption{$p_T(ZH) > 200$}
    \end{subfigure}
    \caption{Transverse Momentum Distribution of $ZH$ }
\end{figure}

This observable provides information on the event
topology and may reveal subtle modifications of the
production mechanism induced by anomalous couplings.

Overall, the kinematic distributions studied in
this analysis illustrate the sensitivity of the
associated Higgs production process to deviations
from the Standard Model predictions.
In the following section, we quantify this sensitivity
by performing a statistical analysis of the
dependence of the production cross section on the
EFT parameter $\epsilon_{ZZ}$.

Figure~4 shows the transverse momentum distribution of the associated $ZH$ system for the Standard Model (SM) prediction and two benchmark scenarios characterized by different values of the anomalous coupling $\epsilon_{ZZ}$.

\paragraph{(a) Full $p_T(ZH)$ spectrum}

The left panel displays the full $p_T(ZH)$ spectrum. As expected for associated Higgs production at hadron colliders, the distribution is strongly peaked at low transverse momentum, where the cross section is dominated by soft and moderately energetic events.
In this region, the predictions corresponding to the Standard Model and the two BSM benchmark points remain very close to each other. This behaviour means that contributions from higher-dimensional operators are suppressed at low energy scales, leading to only minor deviations from the SM prediction. 
Therefore, the low-$p_T$ region shows limited sensitivity to anomalous Higgs couplings.

\paragraph{(b) High-$p_T$ region: $p_T(ZH) > 200~\mathrm{GeV}$}

The right panel focuses on the boosted regime defined by $p_T(ZH) > 200~\mathrm{GeV}$. In this region, clear deviations from the SM prediction emerge.
In particular, the benchmark point with $\epsilon_{ZZ} = +0.5$ exhibits a significant enhancement in the high-$p_T$ tail, leading to a larger number of events compared to the Standard Model expectation. The scenario with $\epsilon_{ZZ} = -0.5$ also shows deviations, although with a smaller magnitude.
This behaviour is characteristic of effective field theory contributions, where the impact of higher-dimensional operators typically grows with the energy scale of the process. As a result, the high-$p_T$ region becomes increasingly sensitive to new physics effects in the Higgs--gauge boson interaction.


These results highlight the importance of boosted Higgs analyses for probing anomalous couplings. While the Standard Model dominates the low-energy regime, the high-$p_T$ tail of the $ZH$ transverse momentum distribution provides a powerful observable to enhance the sensitivity to potential BSM contribution.

\subsection*{Azimuthal Separation}

Finally, we consider the azimuthal angle difference
between the Higgs boson and the $Z$ boson given in equation (4.5).
\begin{figure}[h]
\centering
\includegraphics[width=0.7\textwidth]{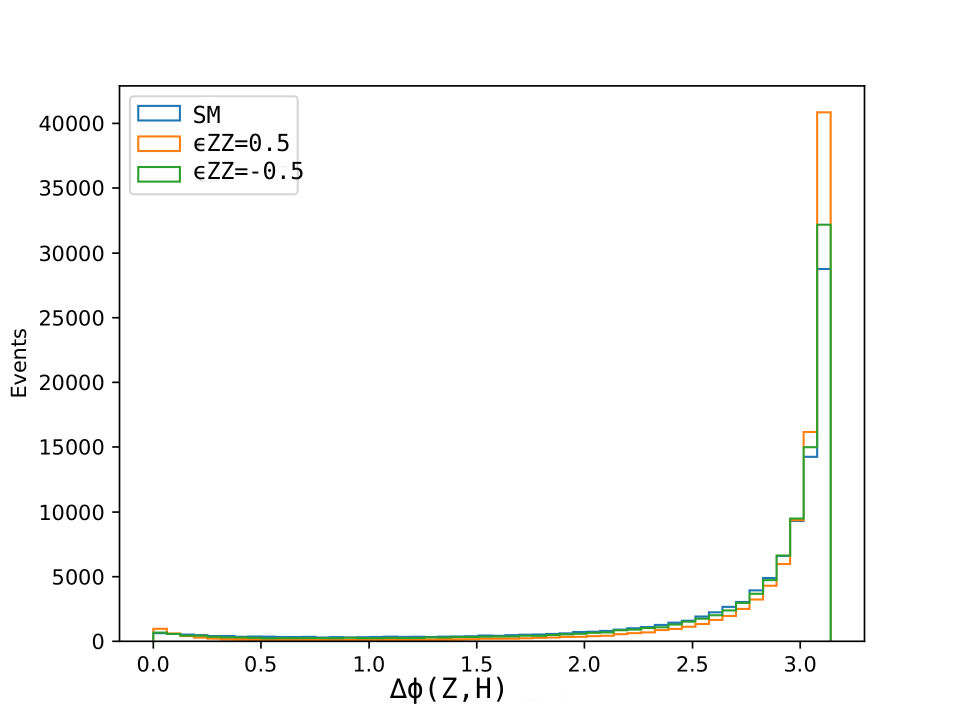}
\caption{Azimuthal separation between the Higgs and
the $Z$ boson.}
\label{fig:dphi}
\end{figure}

This observable provides information on the event
topology and may reveal subtle modifications of the
production mechanism induced by anomalous couplings.

The $\Delta\phi(Z,H)$ distribution exhibits the characteristic back-to-back topology of associated ZH production, with a pronounced peak at $\Delta\phi(Z,H)=\pi$. Variations of the anomalous coupling $\epsilon_{ZZ}$ mainly affect the overall event rate while leaving the shape of the distribution nearly unchanged. This suggests that $\Delta\phi(Z,H)$ has limited shape sensitivity to $\epsilon_{ZZ}$, and the dominant effect arises through modifications of the total ZH production cross section.
$\Delta\phi(Z,H)$ is much less sensitive than $p_{T}H, p_{T}Z, mZH$, for constraining $\epsilon_{ZZ}$.

The results given in Table~\ref{Tab:1}, clearly indicate a strong dependence of the production cross section on the EFT parameter $\epsilon_{ZZ}$, highlighting the sensitivity of the $ZH$ production channel to deviations from the Standard Model prediction.\\
In order to quantify the sensitivity of the associated
Higgs production process to the EFT parameter
$\epsilon_{ZZ}$, a statistical analysis is performed
based on the dependence of the production cross
section on the EFT coupling.

As discussed in the previous section, the total
cross section can be parametrized as a quadratic
function of the EFT parameter:

\begin{equation}
	\sigma(\epsilon_{ZZ}) =
	a + b . \epsilon_{ZZ} + c . \epsilon_{ZZ}^{2}.
\end{equation}

The coefficients of this polynomial are obtained by
fitting the cross sections computed for the three
benchmark points $\epsilon_{ZZ}=0, \epsilon_{ZZ}=0.5$ and
$\epsilon_{ZZ}=-0.5.$

This parametrization allows us to interpolate the
cross section for arbitrary values of the EFT
parameter.

To estimate the allowed range of $\epsilon_{ZZ}$,
a $\chi^{2}$ test is constructed by comparing the
predicted cross section with the Standard Model
expectation

\begin{equation}
\chi^{2}(\epsilon_{ZZ}) =
\frac{
	\left[
	\sigma(\epsilon_{ZZ}) - \sigma_{SM}
	\right]^2
}{
	\delta\sigma^{2}
},
\end{equation}

where $\sigma_{SM}$ is the Standard Model cross
section and $\delta\sigma$ represents the statistical
uncertainty of the simulated sample.

The $\chi^{2}$ function is evaluated for a range of
values of the EFT parameter:

\[
-1 \le \epsilon_{ZZ} \le 1 .
\]

\begin{figure}[h]

\centering

\includegraphics[width=0.7\textwidth]{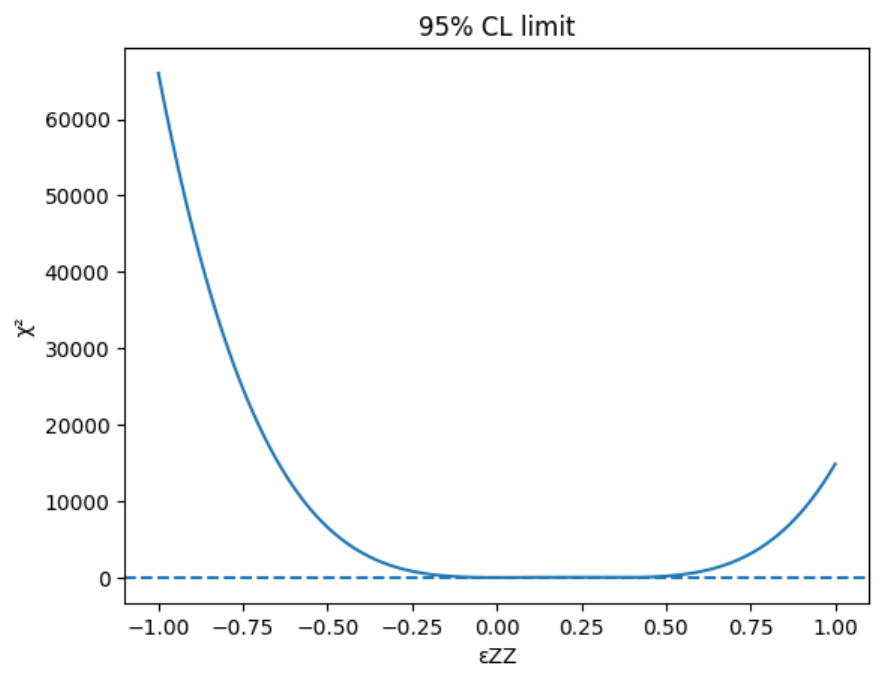}

\caption{$\chi^{2}$ scan as a function of the EFT
parameter $\epsilon_{ZZ}$.}

\label{fig:chi2}

\end{figure}

The resulting $\chi^{2}$ distribution is shown in
Fig.~\ref{fig:chi2}.

The allowed region for the EFT parameter is obtained
by requiring

\[
\chi^{2} < 3.84 ,
\]

which corresponds to a 95\% confidence level
interval for a single parameter.

Applying this criterion to the $\chi^{2}$ scan
yields the following constraint

\[
-0.026 < \epsilon_{ZZ} < 0.383
\quad (95\%~\mathrm{CL}).
\]

This result demonstrates that the associated Higgs
production process provides significant sensitivity
to deviations from the Standard Model Higgs
couplings.

Although the present analysis is based on a
simplified study using simulated event samples,
it illustrates the potential of precision
measurements of the $pp \rightarrow ZH$ process
to constrain anomalous Higgs interactions within
the Effective Field Theory framework.

A more realistic analysis including detector
effects, background contributions, and systematic
uncertainties would further improve the robustness
of the extracted limits.

%% file: sections/Conclusion.tex
\section{Conclusion}

In this work, we have studied the sensitivity of associated
Higgs production, $pp \to ZH$, to modifications of the Higgs
couplings within the Effective Field Theory (EFT) framework,
focusing on the dimension-six operator parametrized by
$\epsilon_{ZZ}$.

We performed a detailed simulation of the signal process
and analyzed the resulting events at the parton level.
The kinematic distributions of the Higgs boson and
associated $Z$ boson were investigated, highlighting
the observables most sensitive to deviations from the
Standard Model.

By performing a $\chi^2$ scan of the total cross section
as a function of $\epsilon_{ZZ}$, we derived a 95\% confidence
level interval

\[
-0.026 < \epsilon_{ZZ} < 0.383 ,
\]

demonstrating that precision measurements of the
$pp \to ZH$ process provide strong constraints on
anomalous Higgs interactions.

Although this study relies on parton-level simulations
and simplified statistical treatment, it illustrates
the potential of associated Higgs production as a
probe of new physics in the Higgs sector.
Inclusion of detector effects, realistic backgrounds,
and systematic uncertainties would be required to
provide limits directly comparable to LHC analyses.

The methodology presented here can be extended to
the High-Luminosity LHC (HL-LHC) and future hadron
colliders, where increased statistics and improved
experimental precision are expected to further enhance
the sensit.

\section*{Funding}
This work was realized with the support of the Algerian Ministry of Higher Education and Scientic Research.

\section*{Conflict of interest}
The authors declare that they have no conflicts of interest.